\documentclass[twocolumn,preprintnumbers,amsmath,amssymb,superscriptaddress,nofootinbib]{revtex4-2}
\usepackage{bm}
\usepackage{braket}
\usepackage{physics}
\usepackage{graphicx}
\usepackage{mathpazo}
 
\newtheorem{theorem}{Theorem}

\begin{document}

\title{Information-geometric bounds on nonequilibrium relaxation in quantum Markov dynamics}

\author{Xiao-Kan Guo}\email{kankuohsiao@whu.edu.cn}
\affiliation{School of Mathematics \& Physics, Yancheng Institute of Technology, Yancheng 224051, Jiangsu, China}
\author{Zhiqiang Huang}\email{zqhuang@hubu.edu.cn}
\affiliation{School of Physics, Hubei University, Wuhan 430062, Hubei, China}

\date{\today}

\begin{abstract}
In this paper, we study the quantum information-geometric structure underlying short-time nonequilibrium relaxation in quantum Markov dynamics, and derive bounds on the nonequilibrium correction to the short-time relaxation curvature. These bounds generalize the information-geometric structure underlying
Auconi's classical nonequilibrium relaxation inequality to quantum Markov dynamics. We consider the von Neumann relative entropy as quantum divergence, and parametrize perturbations through the Kubo-Mori map, which converts the second-order expansion of the relative entropy into the Bogoliubov-Kubo-Mori (BKM) inner product at all temperatures. 
For a quantum Markov semigroup with full-rank stationary state, the induced tangent-space generator admits a canonical decomposition into a BKM-symmetric and a BKM-antisymmetric part, and the leading nonequilibrium curvature correction takes the exact form of the expectation of the commutator between the dissipative and the transport part of the generator. 
This correction is bounded by the product of a BKM-symmetric dissipative activity and a transport-sector activity. 
The bound is verified on a noncommuting qubit model and a driven-dissipative qutrit, and the full chain is confirmed numerically on a two-dimensional Fokker-Planck steady state. 
In the high-temperature and overdamped limit the quantum formula reduces to the position-space Fokker-Planck expression and reproduces Auconi's entropy-production bound with the correct temperature factor. 
\end{abstract}

\maketitle

\section{Introduction}
When a system is driven out of equilibrium, it will in general relax to an equilibrium, or nonequilibrium steady, state due to its internal dynamics or the external influences. Studying  the relaxation processes based on the microscopic dynamics is one of the goals of nonequilibrium statistical mechanics \cite{KTH85}. With different nonequilibrium dynamics, one could obtain very different behaviours of the relaxation time, such as the slow relaxation in critical or glassy systems \cite{Gam08,B15}  and the  anomalous Mpemba effect in which the relaxation time depends  nonmonotonically on the initial distance from the final state \cite{TBLRV26}.

 In spite of the different behaviours of relaxation time, some universal insights can still be learned from the thermodynamic and information-theoretic considerations, cf. \cite{ID20,DGS23,KOI24}.
The relation between relaxation dynamics and thermodynamic dissipation has recently been sharpened by Auconi~\cite{Auconi2025}, who derived an inequality linking the short-time behavior of the Kullback-Leibler divergence to the steady-state entropy production in classical Fokker-Planck systems. Extending such ideas to quantum systems would provide a tool to estimate entropy production in settings where fluctuations are hard to measure. 

In this paper, we carry out  the quantum generalization of nonequilibrium relaxation inequality at short-time scale.
To this end, a necessary first step is to establish a quantum dynamical framework that parallels the classical description. We first generalize the Kullback-Leibler divergence to the quantum relative entropy. For the quantum dynamics, although there are some quantum generalizations of the Fokker-Planck equation \cite{deOliveira2016,deOliveira2020} guaranteeing thermalization to the Gibbs state, we should not confine ourselves to the equilibrium Gibbs state, and relaxation to a nonequilibrium steady state (NESS) is also allowed. We therefore choose to work with the quantum Markov semigroup (QMS)  generated by a Lindblad generator~\cite{Lindblad1976,Breuer2002}.

The classical nonequilibrium relaxation inequality in \cite{Auconi2025} rests on the splitting of the linearized generator into a symmetric dissipative part and an antisymmetric current part. In the quantum generalization, we parametrize perturbations through the Kubo-Mori map, which converts the second-order expansion of the relative entropy into the Bogoliubov-Kubo-Mori (BKM) inner product at all temperatures. The central observation of this paper is that, after pulling the QMS generator back to BKM tangent coordinates, the {\it nonequilibrium correction} to the {\it short-time relaxation curvature} is exactly (at quadratic order in the perturbation amplitude) the expectation of its {\it BKM non-normality}, which identifies generator non-normality as the geometric quantity controlling the departure from detailed-balance relaxation. 
This identification is our new observation, compared with the previous work on  the  Hessian of the relative entropy~\cite{Petz1994}.

For a quantum Markov semigroup with full-rank stationary state, the induced tangent-space generator admits a canonical decomposition \(\mathcal{K}=\mathcal{K}_s+\mathcal{K}_a\) into a BKM-symmetric and a BKM-antisymmetric part, and the leading \(\mathcal{O}(\epsilon^2)\) nonequilibrium curvature correction takes the exact form 
\begin{equation}\xi_Q=\epsilon^2\langle\Phi,[\mathcal{K}_s,\mathcal{K}_a]\Phi\rangle_{\rm BKM},\end{equation}
i.e., the expectation of the commutator between the dissipative and the transport part of the generator---in other words, of the BKM non-normality \([\mathcal{K}^\ddagger,\mathcal{K}]\) of the induced generator. 
 A Cauchy--Schwarz inequality bounds this correction by
\begin{equation}
\xi_Q^2 \le 4\epsilon^4\,\Sigma_s(\Phi)\,\Sigma_a(\Phi).
\end{equation}
with the \emph{dissipative activity of the BKM-symmetric sector}
$
\Sigma_s(\Phi)$,
and the \emph{transport-sector activity}
$
\Sigma_a(\Phi)$.
 
  The bound is verified on a noncommuting qubit model and a driven-dissipative qutrit, and the full chain is confirmed numerically on a two-dimensional Fokker-Planck steady state. In the high-temperature, overdamped limit the quantum formula reduces to the position-space Fokker--Planck expression and reproduces Auconi's entropy-production bound with the correct temperature factor. Two operational perturbation ensembles, random-gradient and displacement, are constructed to estimate the bound without fine-tuning, and the optimality and saturation of the bound are characterized through a universal ratio bound and the associated constrained variational problem. 

This paper is organized as follows.
Section~\ref{sec:prelim} recalls Auconi's classical inequality. Section~\ref{sec:qrelentropy} introduces the quantum relative entropy and its exact second-order expansion. There we parametrize perturbations through the Kubo-Mori map rather than a symmetric ansatz; the expansion then becomes the BKM inner product exactly, at all temperatures. In Section~\ref{sec:tangent}, for a quantum Markov semigroup with full-rank stationary state, we decompose the induced tangent-space generator into its BKM-symmetric (dissipative) and BKM-antisymmetric (transport) parts and prove the quadratic curvature identity together with a Cauchy--Schwarz bound, which we verify on a noncommuting qubit model. In the high-temperature, overdamped reduction the quantum identity reproduces the position-space integral relation and Auconi's entropy-production bound with the correct temperature factor. Section~\ref{sec:ensembles}  constructs operational perturbation ensembles. Section~\ref{sec:numerics} verifies the three levels of the theory numerically on a driven-dissipative qutrit and a two-dimensional Fokker-Planck steady state. Section~\ref{sec:conclusion} concludes.
\section{Classical nonequilibrium relaxation inequality}
\label{sec:prelim}
Consider an overdamped Brownian particle in \(\mathbb{R}^d\) governed by  the stochastic differential equation
\begin{equation}
d\mathbf{x} = \mathbf{F}(\mathbf{x})\,dt + \sqrt{2T}\,d\mathbf{W},\end{equation} 
where $T$ is the temperature, $\mathbf{F}(\mathbf{x})$ is the drift, and $d\mathbf{W}$ is the standard Wiener increment.
The corresponding probability density \(p(\mathbf{x},t)\) satisfies the Fokker-Planck equation \cite{Ito24}
\begin{equation}\partial_t p = -\nabla\cdot(p\bm{\nu})\end{equation}
with \(\bm{\nu} = \mathbf{F} - T\nabla\ln p\). For a NESS \(p^*\) with current \(\bm{\nu}^*\neq0\), the entropy production rate is
 \begin{equation}\sigma(t) = \frac1T\int p\,\|\bm{\nu}\|^2,\end{equation} 
 and its steady-state value is \(\sigma^* = \langle\|\bm{\nu}^*\|^2\rangle/T\) \cite{Seifert2012}, where \(\langle f\rangle\equiv\int p^* f\).

A small perturbation \(p = p^*(1+\phi)\) leads to the linearized dynamics
 \begin{equation}\partial_t\phi = T\nabla^2\phi - (\bm{\nu}^* - T\nabla\ln p^*)\!\cdot\!\nabla\phi.\end{equation} 
 In this linear regime, the Kullback-Leibler divergence \(D(t)=\int p\ln(p/p^*)\) expands to \(D\approx\frac12\langle\phi^2\rangle\), so we have the time derivatives
\begin{align}
\dot D& = -T\langle\|\nabla\phi\|^2\rangle,\nonumber\\
\ddot D &= 2T\bigl[T\langle\alpha^2\rangle - \langle\alpha\,\bm{\nu}^*\!\cdot\!\nabla\phi\rangle\bigr],
\end{align}
where \(\alpha \equiv \nabla\ln p^*\!\cdot\!\nabla\phi + \nabla^2\phi\). Setting \(\bm{\nu}^*=0\) gives the equilibrium value \(\ddot D^{\text{eq}}\). The nonequilibrium correction is defined as
\begin{equation}
\xi \equiv \ddot D - \ddot D^{\text{eq}} = -2T\langle\alpha\,\bm{\nu}^*\!\cdot\!\nabla\phi\rangle.
\label{eq:xi_classical}
\end{equation}
The Cauchy-Schwarz inequality then yields the general bound
\begin{equation}
\sigma^* \ge \frac{\xi^2}{4T^3\,\langle\alpha^2\|\nabla\phi\|^2\rangle},
\label{eq:genbound_classical}
\end{equation}
which can be refined by considering ensembles of perturbations. See \cite{Auconi2025}  for the proof of this inequality.

Two ingredients make the classical bound \eqref{eq:genbound_classical} possible, and the quantum construction below retains these two structural ingredients: (i) the second-order (local) expansion of the divergence around the steady state, and (ii) the splitting of the linearized generator into a symmetric part \(T\nabla^2 + T\nabla\ln p^*\!\cdot\!\nabla\) and an antisymmetric part \(-\bm{\nu}^*\!\cdot\!\nabla\) with respect to the weighted inner product \(\langle f,g\rangle=\int p^* f g\).

\section{Quantum relative entropy and the Kubo--Mori parametrization}
\label{sec:qrelentropy}

The quantum analog of the Kullback--Leibler divergence is the von Neumann relative entropy
\begin{equation}\label{eq:Sdef}
S(\rho\|\rho^*) = \Tr[\rho(\ln\rho - \ln\rho^*)].
\end{equation}
The relative entropy is a divergence, not a metric, as it is asymmetric and does not satisfy the triangle inequality. No metric property will be used below, and only the local quadratic structure of \(S\) around the steady state enters the relaxation inequality. 

We consider a small perturbation around the steady state, \(\rho = \rho^* + \epsilon\,\delta\), with \(\Tr\delta=0\) and \(\delta\) Hermitian. Rather than the symmetric ansatz \(\delta=\frac12\{\rho^*,\Phi\}\), we parametrize \(\delta\) through the Kubo-Mori map \cite{Petz1994}
\begin{equation}
\delta = \Omega_{\rho^*}(\Phi) \equiv \int_0^1 (\rho^*)^s\,\Phi\,(\rho^*)^{1-s}\,ds,
\label{eq:KM_map}
\end{equation}
where \(\Phi\) is Hermitian. The map is invertible, because, in the spectral basis \(\rho^*=\sum_n p_n|n\rangle\langle n|\) it acts as
\begin{equation}
\delta_{mn} = \kappa_{mn}\Phi_{mn},
\qquad
\kappa_{mn} = \frac{p_m-p_n}{\ln p_m-\ln p_n},
\label{eq:kappa}
\end{equation}
with \(\kappa_{nn}=p_n\) and \(\kappa_{mn}=\int_0^1 p_m^s p_n^{1-s}ds\). Since \(\Tr[\Omega_{\rho^*}(\Phi)]=\Tr[\rho^*\Phi]\), the normalization condition \(\Tr\delta=0\) reads \(\Tr[\rho^*\Phi]=0\), i.e. the perturbation has zero steady-state mean. This condition selects the BKM tangent space \(\{\Phi:\Tr[\rho^*\Phi]=0\}\) and can always be enforced by subtracting the steady-state mean, \(\Phi\to\Phi-\Tr[\rho^*\Phi]\). \footnote{The subtraction does not affect any quantity below, because the identity is annihilated by the induced generator, \(\mathcal{K}(1)=\Omega_\pi^{-1}\mathcal{L}(\pi)=0\), hence also \(\mathcal{K}_s(1)=\mathcal{K}_a(1)=0\), so that \(\mathcal{K}\Phi\), \(\mathcal{K}_s\Phi\), \(\mathcal{K}_a\Phi\), and \(\xi_Q\) are unchanged.}

The exact second-order expansion of the relative entropy is (see Appendix~\ref{app:expansion} for the derivation)
\begin{equation}
S(\rho\|\rho^*) = \frac{\epsilon^2}{2}\,
\sum_{m,n} |\delta_{mn}|^2\,\frac{\ln p_m-\ln p_n}{p_m-p_n}
+ \mathcal{O}(\epsilon^3).
\label{eq:S_exact}
\end{equation}
Inserting the parametrization~\eqref{eq:kappa}, we have
\begin{equation}
|\delta_{mn}|^2\,\frac{\ln p_m-\ln p_n}{p_m-p_n}
= |\Phi_{mn}|^2\,\frac{p_m-p_n}{\ln p_m-\ln p_n},
\end{equation}
and therefore
\begin{equation}
S(\rho\|\rho^*) = \frac{\epsilon^2}{2}\,\langle\Phi,\Phi\rangle_{\rho^*} + \mathcal{O}(\epsilon^3)
\label{eq:S_BKM_exact}
\end{equation}
where
\begin{equation}
\langle A,B\rangle_{\rho^*} \equiv \int_0^1 \Tr\!\bigl[\rho^{*s} A^\dagger \rho^{*1-s} B\bigr]\,ds
\label{eq:KM_def}
\end{equation}
is the BKM inner product. This is an \emph{exact} second-order result at arbitrary temperature.
 The quadratic form in~\eqref{eq:S_exact} is the Hessian of the relative entropy at \(\rho^*\), which is precisely the BKM metric of quantum information geometry~\cite{Petz1994}. Although the global divergence is not a metric, its Hessian defines a genuine Riemannian structure, and the relaxation inequality is a short-time, hence local, statement.

In the high-temperature limit the inner product simplifies. As shown in Appendix~\ref{app:Kubo},
\begin{equation}
\langle \Phi,\Phi\rangle_{\rho^*} = \Tr[\rho^* \Phi^\dagger\Phi] + \mathcal{O}((\beta\hbar\omega)^2)
\qquad(\Phi=\Phi^\dagger),
\label{eq:KM_degenerate}
\end{equation}
so that, to leading order,
\begin{equation}
S(\rho\|\rho^*) = \frac{\epsilon^2}{2}\,\Tr[\rho^*\Phi^\dagger\Phi] + \mathcal{O}(\epsilon^3,(\beta\hbar\omega)^2).
\label{eq:S_approx}
\end{equation}
We denote the approximate weighted trace inner product by
\begin{equation}
\langle A,B\rangle \equiv \Tr[\rho^* A^\dagger B].
\label{eq:inner}
\end{equation}
In the same limit the Kubo-Mori map reduces to multiplication by \(\rho^*\), \(\Omega_{\rho^*}(\Phi)\to\rho^*\Phi\).

Assuming the perturbation evolves by the linearized dynamics \(\partial_t\Phi=\mathcal{K}\Phi\) (as in Sec.~\ref{sec:tangent}), the time derivatives of~\eqref{eq:S_BKM_exact} are
\begin{equation}
\frac{dS}{dt} = \epsilon^2\,\Re\langle\Phi,\partial_t\Phi\rangle_{\rho^*} + \mathcal{O}(\epsilon^3),
\label{eq:dS}
\end{equation}
and
\begin{equation}
\frac{d^2S}{dt^2} = \epsilon^2\,\Re\bigl[\langle\partial_t\Phi,\partial_t\Phi\rangle_{\rho^*} + \langle\Phi,\partial_t^2\Phi\rangle_{\rho^*}\bigr] + \mathcal{O}(\epsilon^3).
\label{eq:d2S}
\end{equation}
In the high-temperature reduction the inner product is replaced by the weighted trace~\eqref{eq:inner}.

\section{Tangent-space dynamics and the quantum nonequilibrium correction}
\label{sec:tangent}
Let
\begin{equation}
\partial_t\rho = \mathcal{L}(\rho)
\label{eq:QMS}
\end{equation}
be a QMS generated by a Lindblad generator  $\mathcal{L}$ with a full-rank stationary state \(\pi\), i.e. \(\mathcal{L}(\pi)=0\). No thermal condition is imposed on \(\pi\), so it is the nonequilibrium steady state of the driven dynamics. Writing
\begin{equation}
\rho = \pi + \epsilon\,\Omega_{\pi}(\Phi),
\qquad
\Tr[\pi\Phi]=0,
\label{eq:rho_pi}
\end{equation}
and using the linearity of \(\mathcal{L}\), the perturbation evolves through the induced tangent-space generator
\begin{equation}
\partial_t\Phi = \mathcal{K}\Phi,
\qquad
\mathcal{K} \equiv \Omega_{\pi}^{-1}\,\mathcal{L}\,\Omega_{\pi},
\label{eq:K_def}
\end{equation}
which is well-defined because the spectral kernel \(\kappa_{mn}\) of \(\Omega_\pi\) is strictly positive, and which preserves the zero-mean subspace because \(\mathcal{L}\) is trace preserving.
Throughout this section the stationary state \(\pi\) is full rank and \(\pi>0\), so that \(\Omega_\pi\) is invertible. For finite-dimensional Lindblad dynamics this holds whenever the semigroup is primitive (unique faithful stationary state); stationary states with zero eigenvalues are excluded in what follows.

Let \(\mathcal{K}^{\ddagger}\) denote the adjoint of \(\mathcal{K}\) with respect to the BKM inner product \(\langle\cdot,\cdot\rangle_\pi\), and decompose
\begin{equation}
\mathcal{K} = \mathcal{K}_s + \mathcal{K}_a,
\qquad
\mathcal{K}_s^{\ddagger} = \mathcal{K}_s,
\qquad
\mathcal{K}_a^{\ddagger} = -\mathcal{K}_a.
\label{eq:K_decomp}
\end{equation}
Both \(\mathcal{K}\) and \(\mathcal{K}^\ddagger\) preserve the zero-mean subspace, since \(\mathcal{K}(1)=0\), and \begin{equation}\langle\mathcal{K}^\ddagger\Phi,1\rangle_\pi=\langle\Phi,\mathcal{K}(1)\rangle_\pi=0.\end{equation}
The symmetric part is dissipative: by the data-processing (monotonicity) property of the relative entropy under a QMS, \(\frac{d}{dt}S(\rho_t\|\pi)\le0\), so \begin{equation}\Re\langle\Phi,\mathcal{K}\Phi\rangle_\pi = \langle\Phi,\mathcal{K}_s\Phi\rangle_\pi\le0\end{equation} for all \(\Phi\), i.e., \(\mathcal{K}_s\le0\). The symmetric part defines the \emph{BKM-symmetric reference evolution} \(\partial_t\Phi^{(s)}=\mathcal{K}_s\Phi^{(s)}\) on the tangent space. We emphasize that this is a purely tangent-space reference evolution: as a symmetrization with respect to the BKM inner product, \(\mathcal{K}_s\) need not itself define a completely positive trace-preserving  quantum semigroup, and no physical ``equilibrium channel'' is being assumed. Rather, \(\mathcal{K}_s\) is the BKM-self-adjoint sector associated with the detailed-balance directions of the tangent-space geometry (for a genuine QMS generator, this sector is naturally connected to quantum detailed balance~\cite{FagnolaUmanita2007}). The antisymmetric part \(\mathcal{K}_a\) carries the nonequilibrium current. 

This decomposition is the quantum counterpart of the classical splitting emphasized in Sec.~\ref{sec:prelim}. It has the following transparent geometric meaning. Along any trajectory of~\eqref{eq:K_def},
\begin{equation}
\frac{d}{dt}\,\frac12\|\Phi\|_\pi^2 = \Re\langle\Phi,\mathcal{K}\Phi\rangle_\pi = \langle\Phi,\mathcal{K}_s\Phi\rangle_\pi,
\label{eq:contraction}
\end{equation}
because \(\Re\langle\Phi,\mathcal{K}_a\Phi\rangle_\pi=0\). The symmetric part \(\mathcal{K}_s\) is therefore the \emph{metric-dissipative} component of the generator expressed in BKM tangent coordinates. It controls the contraction of the local information distance. The \emph{metric-antisymmetric} part \(\mathcal{K}_a\) is a transport component that preserves the local BKM norm. We deliberately avoid calling \(\mathcal{K}_a\) the ``irreversible'' part, as its relation to thermodynamic entropy production is not an identity and is established only through the analysis of Sec.~\ref{sec:reduction}. 
Non-normality is usually associated with transient amplification of perturbations. Here, the BKM contraction \eqref{eq:contraction} precludes any transient growth of the local information norm, so the non-normality does not manifest itself at the level of the norm but at the level of its curvature.

\subsection{Curvature identity}
Let \(S^{(s)}(t)\) be the relative entropy along the BKM-symmetric reference evolution \(\Phi^{(s)}(t)=e^{\mathcal{K}_s t}\Phi\), and let \(S(t)\) be the relative entropy along the full dynamics \(\Phi(t)=e^{\mathcal{K}t}\Phi\). \(S^{(s)}\) is thus a geometric reference curvature. From~\eqref{eq:S_BKM_exact} and~\eqref{eq:d2S}, we have
\begin{equation}
\ddot S(0) = \epsilon^2\,\Re\bigl[\langle\mathcal{K}\Phi,\mathcal{K}\Phi\rangle_\pi + \langle\Phi,\mathcal{K}^2\Phi\rangle_\pi\bigr] + \mathcal{O}(\epsilon^3),
\end{equation}
and similarly for \(\ddot S^{(s)}(0)\) with \(\mathcal{K}\to\mathcal{K}_s\). Expanding \(\mathcal{K}=\mathcal{K}_s+\mathcal{K}_a\) and using~\eqref{eq:K_decomp}, the purely antisymmetric terms cancel,
\begin{equation}
\langle\mathcal{K}_a\Phi,\mathcal{K}_a\Phi\rangle_\pi + \langle\Phi,\mathcal{K}_a^2\Phi\rangle_\pi
= \|\mathcal{K}_a\Phi\|_\pi^2 - \|\mathcal{K}_a\Phi\|_\pi^2 = 0,
\end{equation}
while the cross terms combine into 
\begin{multline}
\Re\bigl[\langle\mathcal{K}_s\Phi,\mathcal{K}_a\Phi\rangle_\pi + \langle\mathcal{K}_a\Phi,\mathcal{K}_s\Phi\rangle_\pi\\
+ \langle\Phi,\mathcal{K}_s\mathcal{K}_a\Phi\rangle_\pi + \langle\Phi,\mathcal{K}_a\mathcal{K}_s\Phi\rangle_\pi\bigr]
= 2\Re\langle\mathcal{K}_s\Phi,\mathcal{K}_a\Phi\rangle_\pi .
\end{multline}
Hence
\begin{equation}
\ddot S(0) = 2\epsilon^2\|\mathcal{K}_s\Phi\|_\pi^2 + 2\epsilon^2\,\Re\langle\mathcal{K}_s\Phi,\mathcal{K}_a\Phi\rangle_\pi + \mathcal{O}(\epsilon^3),
\label{eq:ddot_exact}
\end{equation}
\begin{equation}
\ddot S^{(s)}(0) = 2\epsilon^2\|\mathcal{K}_s\Phi\|_\pi^2.
\end{equation}
Both curvatures are known only up to \(\mathcal{O}(\epsilon^3)\). We therefore define the quantum nonequilibrium curvature correction \(\xi_Q\) as the leading \(\mathcal{O}(\epsilon^2)\) contribution to their difference:
\begin{equation}
\xi_Q \equiv \bigl[\ddot S(0) - \ddot S^{(s)}(0)\bigr]_{O(\epsilon^2)}.
\end{equation}
Since \(2\Re\langle\mathcal{K}_s\Phi,\mathcal{K}_a\Phi\rangle_\pi = \langle\Phi,[\mathcal{K}_s,\mathcal{K}_a]\Phi\rangle_\pi\) and \([\mathcal{K}_s,\mathcal{K}_a]=\frac12[\mathcal{K}^\ddagger,\mathcal{K}]\), we obtain 
\begin{align}
\xi_Q &= 2\epsilon^2\,\Re\langle\mathcal{K}_s\Phi,\mathcal{K}_a\Phi\rangle_\pi\nonumber\\
&= \epsilon^2\,\langle\Phi,[\mathcal{K}_s,\mathcal{K}_a]\Phi\rangle_\pi\nonumber\\
&= \frac{\epsilon^2}{2}\,\langle\Phi,[\mathcal{K}^\ddagger,\mathcal{K}]\Phi\rangle_\pi\label{eq:xi_quantum}
\end{align}
The commutator \([\mathcal{K}_s,\mathcal{K}_a]\) is self-adjoint, so \(\xi_Q\) is real.

 Eq.~\eqref{eq:xi_quantum} states that the quadratic short-time nonequilibrium curvature correction is the BKM expectation of the commutator between the metric-dissipative and the metric-antisymmetric parts of the generator, i.e., of the BKM non-normality \([\mathcal{K}^\ddagger,\mathcal{K}]\).
We call \(\xi_Q\) the \emph{BKM non-normality correction}. 

The logical status is worth stating precisely: every detailed-balanced generator (\(\mathcal{K}_a=0\)) is normal and has \(\xi_Q=0\), but normality is strictly weaker than detailed balance, and BKM non-normality is equivalent to the existence of a tangent perturbation with nonzero quadratic curvature correction (the converse follows by polarization applied to the BKM-self-adjoint operator \([\mathcal{K}^\ddagger,\mathcal{K}]\), which annihilates the identity). It should not be conflated with thermodynamic irreversibility.
 Steady-state entropy production and BKM non-normality are thus distinct notions, and the relaxation correction probes the latter.

The generator-level strength of the non-normal coupling is quantified by the numerical radius of this self-adjoint commutator on the physical tangent space \(\mathcal T_\pi=\{\Phi:\Tr[\pi\Phi]=0\}\),
\begin{equation}
\nu_{\rm BKM}(\mathcal{K}) \equiv \sup_{\substack{\Phi\in\mathcal T_\pi\\ \|\Phi\|_\pi=1}}\bigl|\langle\Phi,[\mathcal{K}^\ddagger,\mathcal{K}]\Phi\rangle_\pi\bigr|
= \max_j|\lambda_j([\mathcal{K}^\ddagger,\mathcal{K}])|,
\label{eq:nu_BKM}
\end{equation}
where the last equality holds in finite dimension. \([\mathcal{K}^\ddagger,\mathcal{K}]\) is BKM-self-adjoint and annihilates the identity, so it acts block-diagonally on the BKM-orthogonal decomposition \(\mathbb{C}1\oplus\mathcal T_\pi\), and the restriction to \(\mathcal T_\pi\) does not change the maximal absolute eigenvalue. In terms of \(\nu_{\rm BKM}\),
\begin{equation}
|\xi_Q[\Phi]| \le \frac{\epsilon^2}{2}\,\nu_{\rm BKM}(\mathcal{K})\,\|\Phi\|_\pi^2,
\label{eq:nu_bound}
\end{equation}
for every perturbation \(\Phi\) in the tangent space. 

\subsection{Cauchy-Schwarz bound}
Applying the Cauchy-Schwarz inequality to~\eqref{eq:xi_quantum}, we obtain
\begin{equation}
|\xi_Q| = 2\epsilon^2|\Re\langle\mathcal{K}_s\Phi,\mathcal{K}_a\Phi\rangle_\pi|
\le 2\epsilon^2\|\mathcal{K}_s\Phi\|_\pi\,\|\mathcal{K}_a\Phi\|_\pi,
\end{equation}
so, with the \emph{dissipative activity of the BKM-symmetric sector}
\begin{equation}
\Sigma_s(\Phi) = \langle\mathcal{K}_s\Phi,\mathcal{K}_s\Phi\rangle_\pi,
\label{eq:Sigma_s}
\end{equation}
and the \emph{transport-sector activity}
\begin{equation}
\Sigma_a(\Phi) = \langle\mathcal{K}_a\Phi,\mathcal{K}_a\Phi\rangle_\pi,
\label{eq:Sigma_a}
\end{equation}
we obtain the following.
\begin{theorem}[Activity-current bound]
\label{thm:bound}
For every perturbation \(\Phi\) with \(\Tr[\pi\Phi]=0\),
\begin{equation}
\xi_Q^2 \le 4\epsilon^4\,\Sigma_s(\Phi)\,\Sigma_a(\Phi).
\label{eq:quantum_CS}
\end{equation}
Equality holds if and only if \(\mathcal{K}_s\Phi\) and \(\mathcal{K}_a\Phi\) are collinear in the BKM inner product, \(\mathcal{K}_s\Phi=c\,\mathcal{K}_a\Phi\) with \(c\in\mathbb{R}\) (possibly with one side vanishing); if no nontrivial collinear pair exists the bound is strict and can only be approached along aligning sequences.
\end{theorem}
The symmetric-sector activity is directly determined by the relaxation curvature under the reference evolution, \(\ddot S^{(s)}(0)=2\epsilon^2\Sigma_s(\Phi)\), so the inequality bounds the transport activity by relaxation data. In the semiclassical reduction of Sec.~\ref{sec:reduction}, \(\Sigma_a\) is bounded by the steady-state entropy production, which turns~\eqref{eq:quantum_CS} into the explicit bound of Eq.~\eqref{eq:quantumbound}.

As an immediate example, let us consider a noncommuting quantum Markov generator.
Take the generator
\begin{equation}
\mathcal{L}(\rho) = -i\Omega[\sigma_x,\rho] + \gamma(\sigma_z\rho\sigma_z - \rho),
\label{eq:qubit_L}
\end{equation}
whose stationary state is the maximally mixed state \(\pi=I/2\), with \(\Omega_\pi(\Phi)=\Phi/2\) and \(\langle X,Y\rangle_\pi=\frac12\Tr[X^\dagger Y]\). This example demonstrates the genuinely noncommutative content of the curvature identity; it is not meant to represent a finite-temperature nonequilibrium steady state, since its stationary state is the maximally mixed state, whose Spohn entropy production vanishes \cite{Spohn1978}. Then
\begin{equation}
\mathcal{K}_s\Phi = \gamma(\sigma_z\Phi\sigma_z-\Phi),
\qquad
\mathcal{K}_a\Phi = -i\Omega[\sigma_x,\Phi].
\end{equation}
For \(\Phi = a\sigma_x+b\sigma_y+c\sigma_z\) with \(a,b,c\in\mathbb{R}\),
\begin{equation}
\mathcal{K}_s\Phi = -2\gamma(a\sigma_x+b\sigma_y),
\qquad
\mathcal{K}_a\Phi = 2\Omega(b\sigma_z-c\sigma_y),
\end{equation}
and a direct computation gives
\begin{equation}
\xi_Q = 8\epsilon^2\gamma\Omega\,bc,
\quad
\Sigma_s = 4\gamma^2(a^2+b^2),
\quad
\Sigma_a = 4\Omega^2(b^2+c^2).
\label{eq:qubit_xi}
\end{equation}
For \(\Phi=\sigma_y+\sigma_z\), \(\xi_Q=8\epsilon^2\gamma\Omega\), \(\Sigma_s=4\gamma^2\), \(\Sigma_a=8\Omega^2\), and~\eqref{eq:quantum_CS} is verified.

 Three features deserve emphasis. (i) The dynamics is genuinely noncommutative, as the Hamiltonian and dissipator do not commute, \([H,L_\mu]=\Omega[\sigma_x,\sigma_z]=-2i\Omega\,\sigma_y\neq0\), so the process is not a classical rate equation, and \(\xi_Q\neq0\) requires a perturbation that is rotated by the Hamiltonian (\(b,c\neq0\)). (ii) In the Hamiltonian-free limit \(\Omega\to0\) the BKM-antisymmetric sector vanishes, \(\mathcal{K}_a\to0\), and \(\xi_Q\to0\), as required. (iii) The dimensionless ratio
\begin{equation}
\frac{\xi_Q^2}{4\epsilon^4\Sigma_s\Sigma_a}
= \frac{b^2c^2}{(a^2+b^2)(b^2+c^2)} \le 1
\label{eq:qubit_ratio}
\end{equation}
has supremum \(1\), approached along the family \((a,b,c)=(0,1,n)\), as \(n\to\infty\), because the collinearity condition \(\mathcal{K}_s\Phi\propto\mathcal{K}_a\Phi\) has no nontrivial solution in this model. Since \(\xi_Q\propto\gamma\Omega\), \(\Sigma_s\propto\gamma^2\), and \(\Sigma_a\propto\Omega^2\), the normalized ratio~\eqref{eq:qubit_ratio} is independent of the rates and depends only on the direction of \(\Phi\) in the Pauli basis.

\subsection{Tightness and optimization}
\label{sec:tightness}
A useful universal measure of tightness is the ratio bound obtained as in \eqref{eq:qubit_ratio},
\begin{equation}
R[\Phi] \equiv \frac{\xi_Q^2}{4\epsilon^4\Sigma_s(\Phi)\Sigma_a(\Phi)}
= \frac{|\Re\langle\mathcal{K}_s\Phi,\mathcal{K}_a\Phi\rangle_\pi|^2}{\|\mathcal{K}_s\Phi\|_\pi^2\,\|\mathcal{K}_a\Phi\|_\pi^2} \le 1,
\label{eq:ratio_bound}
\end{equation}
with equality if and only if \(\mathcal{K}_s\Phi\) and \(\mathcal{K}_a\Phi\) are collinear, \(\mathcal{K}_s\Phi = c\,\mathcal{K}_a\Phi\) with \(c\in\mathbb{R}\). Equivalently, defining the BKM sector-alignment angle by
\begin{equation}
\cos\theta_\Phi \equiv \frac{\Re\langle\mathcal{K}_s\Phi,\mathcal{K}_a\Phi\rangle_\pi}{\|\mathcal{K}_s\Phi\|_\pi\,\|\mathcal{K}_a\Phi\|_\pi}.
\label{eq:cos_angle}
\end{equation}
 Accordingly we interpret \(R[\Phi]=\cos^2\theta_\Phi\in[0,1]\) as the \emph{visibility} of the BKM non-normal coupling in the perturbation \(\Phi\) (with the convention \(R[\Phi]=0\) whenever \(\Sigma_s(\Phi)\Sigma_a(\Phi)=0\), in which case \(\xi_Q[\Phi]=0\)). 
 It measures how strongly a given perturbation couples the dissipative and transport sectors of the generator, with \(R=0\) when the two sectors do not interfere on \(\Phi\), and \(R\to1\) when the dissipative and transport responses become maximally aligned.
 
 At fixed sector activities, the optimal perturbation is the one that maximizes the visibility \(R[\Phi]\), i.e., the perturbation that maximally probes the non-normal sector of the generator. If one instead fixes the two scales, \(\langle\Phi,\mathcal{K}_s^2\Phi\rangle_\pi=1\) and \(-\langle\Phi,\mathcal{K}_a^2\Phi\rangle_\pi=1\), the stationarity condition of the constrained maximization of \(|\langle\Phi,[\mathcal{K}_s,\mathcal{K}_a]\Phi\rangle_\pi|\) is the two-constraint generalized eigenvalue problem
\begin{equation}
[\mathcal{K}_s,\mathcal{K}_a]\,\Phi = \lambda_1\,\mathcal{K}_s^2\,\Phi - \lambda_2\,\mathcal{K}_a^2\,\Phi,
\label{eq:genev}
\end{equation}
with \(\langle\Phi,[\mathcal{K}_s,\mathcal{K}_a]\Phi\rangle_\pi = \lambda_1+\lambda_2\le2\) at the constrained optimum. This is an optional variational formulation for computing the best perturbation at fixed symmetric-sector activity and transport activity.

\section{High-temperature, overdamped reduction}
\label{sec:reduction}

We now show that the quadratic curvature relation~\eqref{eq:xi_quantum} reduces to Auconi's classical expression in the high-temperature, overdamped limit. In this limit the operators become classical functions of position, \(\Omega_{\rho^*}(\Phi)\to\rho^*\Phi\), \(\langle\cdot,\cdot\rangle_{\rho^*}\to\langle\cdot,\cdot\rangle\) of~\eqref{eq:inner}, and the quantum dynamics reduces to the position-space Fokker-Planck equation for the deviation \(\delta=\rho-\rho^*\):
\begin{equation}
\partial_t \delta = -\nabla\!\cdot\!(\mathbf{F}^* \delta) + T\nabla^2\delta,
\label{eq:classical_linear}
\end{equation}
where \(\mathbf{F}^*(\mathbf{x})\) is the effective drift. The steady-state condition is \(-\nabla\!\cdot\!(\mathbf{F}^*\rho^*) + T\nabla^2\rho^* = 0\), and the steady-state probability current
\begin{equation}
\mathbf{J}^* = \mathbf{F}^*\rho^* - T\nabla\rho^*
\label{eq:Jstar}
\end{equation}
is nonvanishing in the NESS. The drift decomposes as
\begin{equation}
\mathbf{F}^* = T\nabla\ln\rho^* + \mathbf{v}^*,
\qquad
\mathbf{v}^* = \frac{\mathbf{J}^*}{\rho^*},
\qquad
\nabla\!\cdot\!(\rho^*\mathbf{v}^*)=0,
\label{eq:F_split}
\end{equation}
where the first term is the detailed-balanced (equilibrium) part and \(\mathbf{v}^*\) the current velocity. Defining \(\delta = \rho^*\Phi\) and using \(\nabla\!\cdot\!\mathbf{J}^*=0\), a straightforward expansion (Appendix~\ref{app:linearize}) yields the linearized equation for \(\Phi\):
\begin{equation}
\partial_t \Phi = \frac{1}{\rho^*}T\nabla\!\cdot\!(\rho^*\nabla\Phi) - \mathbf{v}^*\!\cdot\!\nabla\Phi
\label{eq:Phi_lin_correct}
\end{equation}
The first term is the equilibrium Fokker--Planck generator
\begin{equation}
\mathcal{L}^{\text{eq}}_\Phi = \frac{1}{\rho^*}T\nabla\!\cdot\!(\rho^*\nabla) = T\nabla^2 + T\nabla\ln\rho^*\!\cdot\!\nabla,
\label{eq:Leq_def}
\end{equation}
which is self-adjoint with respect to the weighted inner product~\eqref{eq:inner},
\begin{equation}
\langle A,\mathcal{L}^{\text{eq}}_\Phi B\rangle = -T\langle\nabla A,\nabla B\rangle,
\label{eq:db}
\end{equation}
and the second term \(-\mathbf{v}^*\!\cdot\!\nabla\) is antisymmetric, since \(\nabla\!\cdot\!(\rho^*\mathbf{v}^*)=0\) implies
\begin{equation}
\langle A,\mathbf{v}^*\!\cdot\!\nabla B\rangle = -\langle\mathbf{v}^*\!\cdot\!\nabla A,B\rangle.
\end{equation}
The induced generator is therefore
\begin{equation}
\partial_t \Phi = \mathcal{L}^{\text{eq}}_\Phi \Phi - \mathbf{v}^*\!\cdot\!\nabla\Phi,
\label{eq:Phi_lin_final}
\end{equation}
which is precisely the reduction of the decomposition~\eqref{eq:K_decomp}:
\begin{equation}
\mathcal{K}_s \;\to\; \mathcal{L}^{\text{eq}}_\Phi,
\qquad
\mathcal{K}_a \;\to\; -\mathbf{v}^*\!\cdot\!\nabla.
\label{eq:K_reduction}
\end{equation}

Inserting~\eqref{eq:K_reduction} into the quadratic curvature identity~\eqref{eq:xi_quantum} and using~\eqref{eq:db} with integration by parts,
\begin{equation}
\xi_Q = 2\epsilon^2\,\Re\langle\mathcal{L}^{\text{eq}}_\Phi\Phi,\,-\mathbf{v}^*\!\cdot\!\nabla\Phi\rangle
= -2\epsilon^2 T\,\langle\alpha,\mathbf{v}^*\!\cdot\!\nabla\Phi\rangle,
\label{eq:xiQ_final_scalar}
\end{equation}
where
\begin{equation}
\alpha \equiv \nabla\ln\rho^*\!\cdot\!\nabla\Phi + \nabla^2\Phi.
\label{eq:alpha_def}
\end{equation}
Defining the vector operator
\begin{equation}
\mathcal{A}(\Phi) \equiv \alpha\,\nabla\Phi,
\label{eq:def_A}
\end{equation}
we can write the result compactly as
\begin{equation}
\xi_Q = -2\epsilon^2 T\,\langle \mathbf{v}^*, \mathcal{A}(\Phi) \rangle.
\label{eq:xiQ_compact}
\end{equation}
Here, for vector operators \(\mathbf{a}, \mathbf{b}\), the inner product is understood as \(\langle \mathbf{a},\mathbf{b}\rangle = \Tr[\rho^*\,\mathbf{a}^\dagger\!\cdot\!\mathbf{b}]\).

 Eq. \eqref{eq:xiQ_compact} is the semiclassical form of the quantum curvature relation. In the classical limit it reduces exactly to Auconi's expression \eqref{eq:xi_classical}, since \(\alpha\) and \(\nabla\Phi\) commute and \(\mathbf{v}^*\) becomes the classical velocity field \(\bm{\nu}^*\). Note that the factor \(\epsilon^2\) appears explicitly because the perturbation amplitude is measured by \(\epsilon\).

In the same limit, the entropy production rate reduces to the standard 
\begin{equation}
\sigma_Q^* = \frac1T\,\Tr\!\bigl[\rho^*\,\|\mathbf{v}^*\|^2\bigr] = \frac1T\,\langle\mathbf{v}^*,\mathbf{v}^*\rangle.
\label{eq:sigma_def}
\end{equation}
Applying the weighted Cauchy--Schwarz inequality to \eqref{eq:xiQ_compact} yields
\begin{equation}
\xi_Q^2 = 4\epsilon^4 T^2 \,|\langle\mathbf{v}^*,\mathcal{A}(\Phi)\rangle|^2 \le 4\epsilon^4 T^2\,\langle\mathbf{v}^*,\mathbf{v}^*\rangle\,\langle\mathcal{A}(\Phi),\mathcal{A}(\Phi)\rangle.
\label{eq:CS}
\end{equation}
Define
\begin{equation}
\Sigma(\Phi) \equiv \langle\mathcal{A}(\Phi),\mathcal{A}(\Phi)\rangle = \Tr[\rho^*\,\alpha^2\,\|\nabla\Phi\|^2],
\label{eq:Sigma_def}
\end{equation}
which depends only on the perturbation and the steady state. Using \eqref{eq:sigma_def}, we obtain
\begin{equation}
\sigma_Q^* \ge \frac{\xi_Q^2}{4\epsilon^4 T^3\,\Sigma(\Phi)}.
\label{eq:quantumbound}
\end{equation}
This is the semiclassical reduction of the general bound \eqref{eq:quantum_CS}. Indeed, \(\Sigma_s(\Phi)\to T^2\langle\alpha^2\rangle\) and \(\Sigma_a(\Phi)\to\langle(\mathbf{v}^*\!\cdot\!\nabla\Phi)^2\rangle\), so that \eqref{eq:quantum_CS} gives \(\xi_Q^2 \le 4\epsilon^4T^2\langle\alpha^2\rangle\langle(\mathbf{v}^*\!\cdot\!\nabla\Phi)^2\rangle\).

At the level of the reduced theory, \begin{equation}\Sigma_a(\Phi)=\langle(\mathbf{v}^*\!\cdot\!\nabla\Phi)^2\rangle \le T\,\|\nabla\Phi\|_\infty^2\,\sigma_Q^*\end{equation} which turns~\eqref{eq:quantum_CS} into the (weaker) bound \begin{equation}\sigma_Q^* \ge \xi_Q^2/(4\epsilon^4 T\Sigma_s(\Phi)\|\nabla\Phi\|_\infty^2);\end{equation} the componentwise version~\eqref{eq:CS} is sharper, since \(\langle\alpha^2\|\nabla\Phi\|^2\rangle \le \langle\alpha^2\rangle\|\nabla\Phi\|_\infty^2\). Whether an analogous generator-level bound \(\Sigma_a(\Phi)\le C(\Phi)\,\sigma_{\rm ss}\) holds for a suitable quantum definition of the steady-state entropy production \(\sigma_{\rm ss}\) is the main open problem.
\section{Constructing operational perturbation ensembles}
\label{sec:ensembles}

To apply the bound without fine-tuning, we construct two families of perturbations, normalized to \(O(1)\) amplitude. The small parameter \(\epsilon\) appearing in the definitions below is the same as the expansion parameter used in the density operator expansion \(\rho = \rho^* + \epsilon\delta\).

\subsection{Random-gradient ensemble}
We prepare perturbations of the form
\begin{equation}
\Phi_{\theta,\psi}(\mathbf{x}) = \sin(k\,\mathbf{e}_\theta\!\cdot\!\mathbf{x} + \psi) - \Tr[\rho^*\sin(k\,\mathbf{e}_\theta\!\cdot\!\mathbf{x} + \psi)],
\end{equation}
with \(k\) small, \(\mathbf{e}_\theta=(\cos\theta,\sin\theta)\), and \(\theta,\psi\) uniformly distributed. The mean subtraction enforces the tangent condition \(\Tr[\rho^*\Phi]=0\) for every realization, and it leaves \(\nabla\Phi\), \(\alpha\), \(\mathcal{A}(\Phi)\), and every quantity entering the bound unchanged, since these depend only on gradients of \(\Phi\). The gradient \(\nabla\Phi = k\,\mathbf{e}_\theta\cos(k\,\mathbf{e}_\theta\!\cdot\!\mathbf{x}+\psi)\) is an approximately constant vector with random direction and phase. With \(u=k\,\mathbf{e}_\theta\!\cdot\!\mathbf{x}+\psi\), the exact computation gives \(\alpha = k\,(\mathbf{e}_\theta\!\cdot\!\nabla\ln\rho^*)\cos u - k^2\sin u\), hence
\begin{equation}
\mathcal{A}(\Phi_{\theta,\psi}) = \alpha\nabla\Phi = k^2\,\mathcal{B}_\theta\cos^2 u - k^3\,\mathbf{e}_\theta\sin u\cos u,
\end{equation}
with \(\mathcal{B}_\theta = (\mathbf{e}_\theta\!\cdot\!\nabla\ln\rho^*)\,\mathbf{e}_\theta\), and therefore
\begin{multline}
\xi_Q(\theta,\psi) = -2\epsilon^2 T k^2\,\langle\mathbf{v}^*,\mathcal{B}_\theta\rangle\cos^2 u\\
+ 2\epsilon^2 T k^3\,\langle\mathbf{v}^*,\mathbf{e}_\theta\rangle\sin u\cos u.
\end{multline}
Note that the ensemble is not phase-centered: since \(\mathbb{E}_\psi[\cos^2 u]=1/2\), the leading phase average is \(\mathbb{E}_\psi[\xi_Q] = -\epsilon^2 T k^2\,\langle\mathbf{v}^*,\mathcal{B}_\theta\rangle\), generically nonzero. Using \(\mathbb{E}_\psi[\cos^4 u]=3/8\), \(\mathbb{E}_\psi[\sin u\cos^3 u]=0\), and \(\mathbb{E}_\psi[\sin^2 u\cos^2 u]=1/8\), the second moment is
\begin{equation}
\mathbb{E}_\psi[\xi_Q^2] = \frac32\,\epsilon^4 T^2 k^4\,|\langle\mathbf{v}^*,\mathcal{B}_\theta\rangle|^2 + \mathcal{O}(k^6),
\end{equation}
while \(\mathbb{E}_\psi[\Sigma(\Phi)] = \frac38 k^4\,\langle\mathcal{B}_\theta,\mathcal{B}_\theta\rangle + \mathcal{O}(k^6)\). Combining the pointwise bound \eqref{eq:quantumbound}, written as \(\xi_Q^2 \le 4\epsilon^4T^3\sigma_Q^*\Sigma(\Phi)\), with these averages over \(\psi\) and \(\theta\), we obtain the ensemble bound
\begin{equation}
\sigma_Q^* \ge \frac{\mathbb{E}_{\theta,\psi}[\xi_Q^2]}{4\epsilon^4 T^3\,\mathbb{E}_{\theta,\psi}[\Sigma(\Phi)]}
= \frac{\mathbb{E}_\theta\bigl[|\langle\mathbf{v}^*,\mathcal{B}_\theta\rangle|^2\bigr]}{T\,\mathcal{J}_F} + \mathcal{O}(k^2),
\label{eq:ensemble_bound}
\end{equation}
where \(\mathcal{J}_F = \mathbb{E}_\theta\langle\mathcal{B}_\theta,\mathcal{B}_\theta\rangle = \frac12\Tr[\rho^*\|\nabla\ln\rho^*\|^2]\) (in two dimensions). A centered variant, obtained by subtracting the computable phase mean \(-\epsilon^2Tk^2\langle\mathbf{v}^*,\mathcal{B}_\theta\rangle\), leads to a bound of the same form with a prefactor three times smaller.

\subsection{Displacement ensemble}
A uniform spatial displacement is generated by \(U_\theta(\epsilon) = e^{-i\epsilon\mathbf{e}_\theta\cdot\mathbf{P}}\). To first order in \(\epsilon\), the perturbed state corresponds to
\begin{equation}
\Phi_\theta(\mathbf{x}) = -\mathbf{e}_\theta\!\cdot\!\nabla\ln\rho^*.
\end{equation}
Computing \(\alpha\) and \(\mathcal{A}(\Phi_\theta)\) and defining \(\mathcal{R} = \mathbb{E}_\theta\langle\mathcal{A}(\Phi_\theta),\mathcal{A}(\Phi_\theta)\rangle\), we find
\begin{equation}
\sigma_Q^* \ge \frac{(\mathbb{E}[\xi_Q])^2}{4\epsilon^4 T^3\mathcal{R}}.
\label{eq:displacement_bound}
\end{equation}
Both inequalities relate the steady-state entropy production to short-time relaxation data and properties of the stationary state.

\section{Numerical verification}
\label{sec:numerics}

We verify the three levels of the theory, i.e. the exact quadratic curvature identity, the non-normality/visibility structure, and the semiclassical entropy-production bound, on a finite-dimensional quantum NESS and on a two-dimensional Fokker-Planck NESS. The identity, the activity bound, and the visibility structure are exact statements about an arbitrary QMS, whereas the entropy-production bound \eqref{eq:quantumbound} is verified here in its semiclassical regime only.

\subsection{Driven-dissipative qutrit.}
Consider the three-level GKLS generator
\begin{equation}
\mathcal{L}(\rho) = -i[H,\rho] + \sum_{j\neq k}\gamma_{j\to k}\bigl(|k\rangle\langle j|\,\rho\,|j\rangle\langle k| - \tfrac12\{|j\rangle\langle j|,\rho\}\bigr),
\label{eq:qutrit_gk}
\end{equation}
with the Hamiltonian
\begin{multline}
H=\omega_1|1\rangle\langle1|+\omega_2|2\rangle\langle2|\\
+g(|0\rangle\langle1|+|1\rangle\langle0|+|1\rangle\langle2|+|2\rangle\langle1|),
\end{multline}
the coupling parameters \((\omega_1,\omega_2,g)=(1.0,1.7,0.8)\), and the biased transition rates \((\gamma_{0\to1},\gamma_{1\to0},\gamma_{1\to2},\gamma_{2\to1})=(2.0,0.3,1.1,0.6)\). The stationary state is full rank, with spectrum \(\mathrm{spec}(\pi)=\{0.0875,0.3636,0.5488\}\), with \([H,L_\mu]\neq0\), and \(\nu_{\rm BKM}(\mathcal{K})=4.56\).

The correction \(\xi_Q\) is evaluated in two independent ways: algebraically, as \(\epsilon^2\langle\Phi,[\mathcal{K}_s,\mathcal{K}_a]\Phi\rangle_\pi\), and directly from the definition, by extracting the leading \(\mathcal{O}(\epsilon^2)\) contribution of \(\ddot S(0)-\ddot S^{(s)}(0)\) through a complex-step mixed fourth derivative of \(S(\pi+\epsilon\,\Omega_\pi(e^{\mathcal{K}t}\Phi)\|\pi)\) in \((\epsilon,t)\), with the finite-dimensional matrix-function form of the relative entropy analytically continued to complex \(\epsilon\) solely for derivative evaluation, without using the \(K_s/K_a\) decomposition in the direct extraction. Over \(120\) random BKM-normalized, zero-mean Hermitian perturbations, the two agree to within \(5\times10^{-4}\) relative error [Fig.~\ref{fig:num_identity}]. The extraction is stable under an \(\epsilon\)-scan: the deviation of \(\xi_Q(\epsilon)/\epsilon^2\) from its limiting value is linear in \(\epsilon\) (the deviations at \(\epsilon=10^{-2},\,5\times10^{-3},\,2.5\times10^{-3}\) stand in the ratio \(4.2:2.1:1\)), consistent with the expected \(\mathcal{O}(\epsilon^3)\) correction to \(\xi_Q(\epsilon)\) itself. We next test whether the generator-level non-normality is itself sufficient to produce a large relaxation signal. Theorem~\ref{thm:bound} and the bound \eqref{eq:nu_bound} are confirmed (\(R[\Phi]\le0.067\) and \(Q[\Phi]\le0.64\) over the ensemble), where \(Q[\Phi]\equiv 2|\xi_Q|/(\epsilon^2\nu_{\rm BKM}\|\Phi\|_\pi^2)\le1\) is the normalized ratio of the generator-level bound, while \(R[\Phi]=\cos^2\theta_\Phi\) is the alignment visibility. The more informative statement is the following. Along the one-parameter family \(\Phi(\theta)=\cos\theta\,\Phi_+ + \sin\theta\,\Phi_-\) connecting the two eigendirections of \([\mathcal{K}^\ddagger,\mathcal{K}]\) with eigenvalues \(\pm\nu_{\rm BKM}\), the normalized ratio \(Q[\Phi]\) sweeps the interval \([0.013,0.9999]\) at \emph{fixed} \(\nu_{\rm BKM}\) [Fig.~\ref{fig:num_visibility}]: generator-level non-normality is constant along the family, while the observable signal varies by two orders of magnitude. The contrast with the random ensemble (max \(R=0.067\)) shows that the sampled random perturbations exhibit weak visibility, whereas a properly aligned perturbation nearly saturates the bound (\(Q\simeq0.9999\)). Large generator non-normality does not by itself imply a large observable curvature, which is controlled by the sector alignment \(\cos\theta_\Phi\).

\subsection{Two-dimensional Fokker-Planck NESS.}
The reduced case of Secs.~\ref{sec:reduction} is tested on \begin{equation}\partial_t\rho=-\nabla\cdot(\mathbf{F}\rho)+T\nabla^2\rho\end{equation} with \(T=1\) and \begin{equation}\mathbf{F}=-\nabla V+\Omega_d(-y,x), ~V=\frac12(\omega_x^2x^2+\omega_y^2y^2).\end{equation} 
Throughout this subsection \(\sigma^*\) denotes the steady entropy production of this Fokker-Planck NESS, i.e., the quantity \(\sigma_Q^*\) entering the bound \eqref{eq:quantumbound}. As a control, the isotropic case \(\omega_x=\omega_y=1\) is exactly solvable (\(\rho^*\propto e^{-(x^2+y^2)/2T}\), \(\mathbf{v}^*=\Omega_d(-y,x)\), \(\sigma^*=2\Omega_d^2/\omega_x^2=0.5\); the numerics give \(\sigma^*=0.499\)): it supports a finite stationary current and positive entropy production, but \([\mathcal{K}_s,\mathcal{K}_a]\propto[\mathcal{L}^{\rm eq}_\Phi,\partial_\varphi]=0\), so \(\xi_Q\equiv0\), and numerically its maximal curvature correction (\(0.010\)) is three orders of magnitude below the anisotropic values below. Entropy production and BKM non-normality are thus not interchangeable diagnostics.

\begin{figure}[t]
\includegraphics[width=\columnwidth]{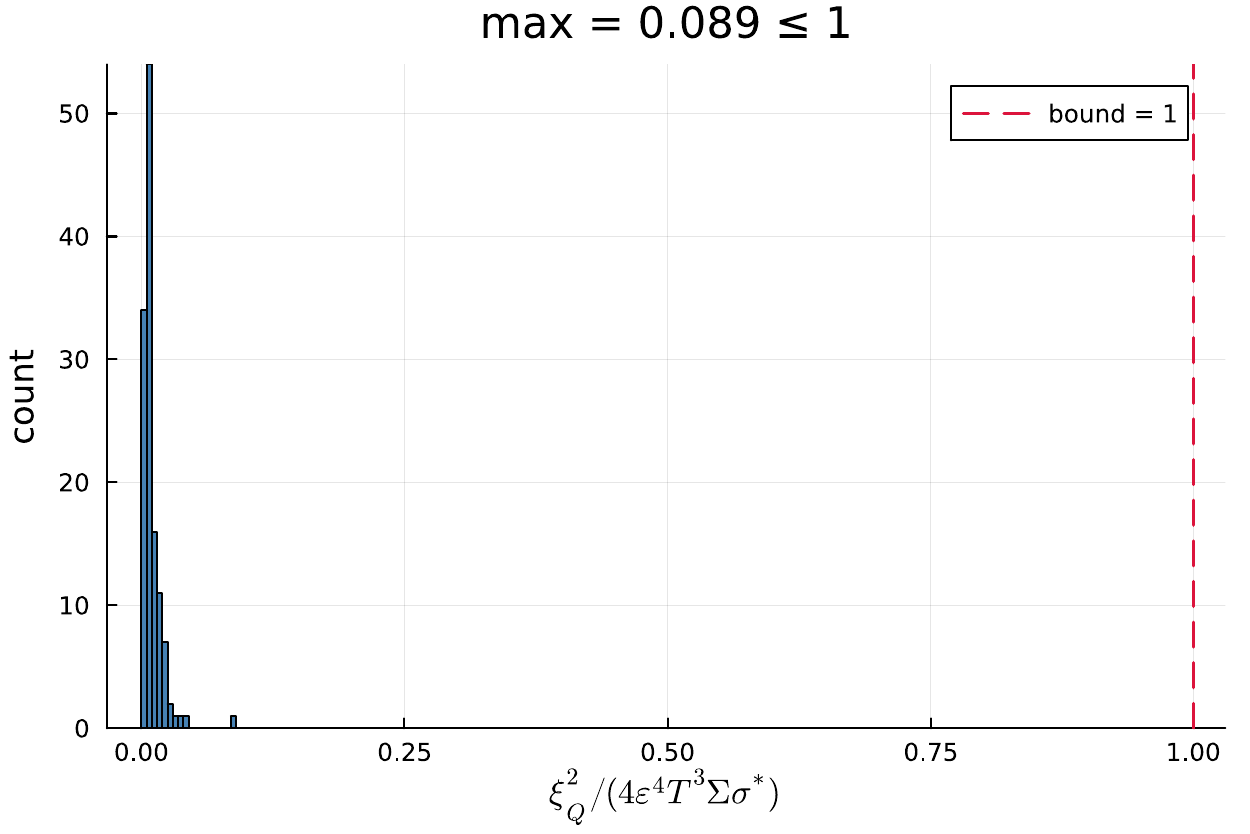}
\caption{Semiclassical entropy-production bound: histogram of \(\xi_Q^2/(4\epsilon^4T^3\Sigma\,\sigma^*)\) over \(128\) perturbations of the two-dimensional anisotropic NESS; all values lie below \(1\) (maximum \(0.089\)).}
\label{fig:num_bound}
\end{figure}

\begin{figure}[t]
\includegraphics[width=\columnwidth]{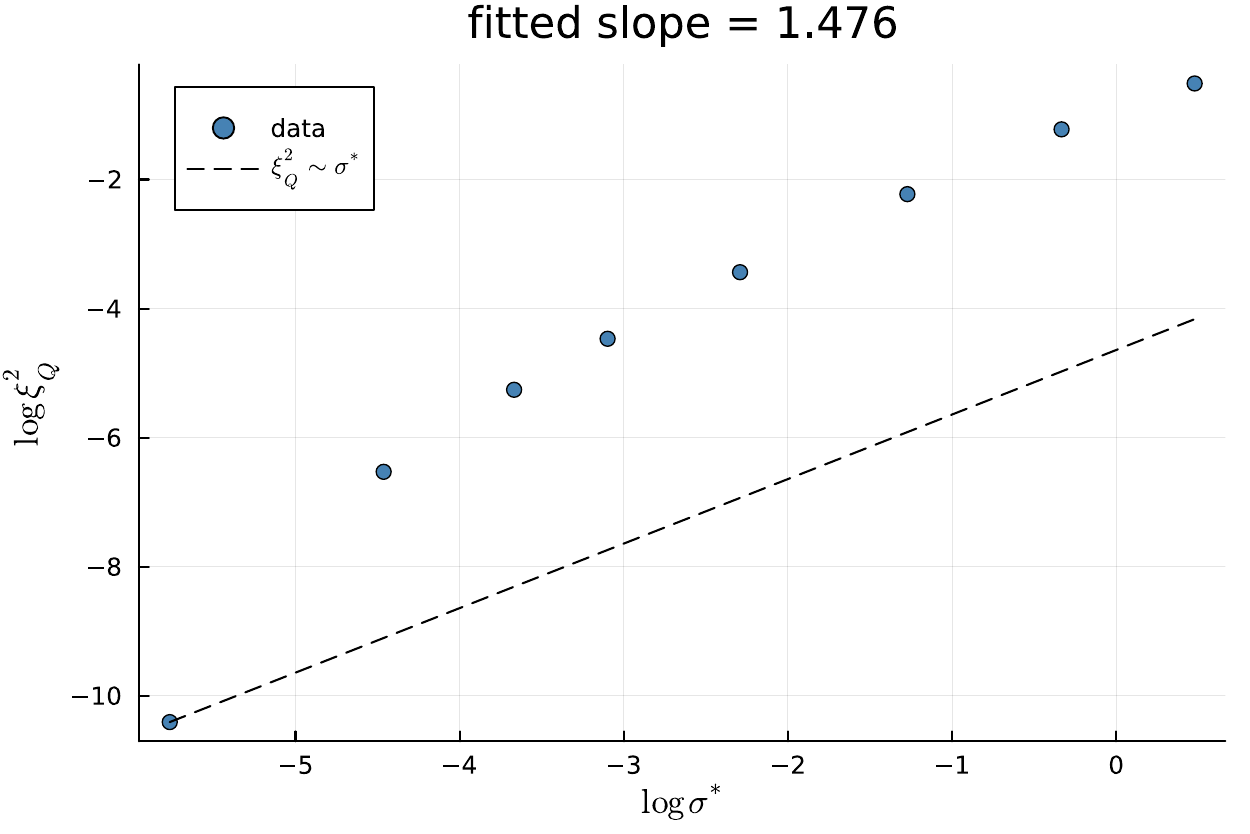}
\caption{Drive scaling: \(\log\xi_Q^2\) versus \(\log\sigma^*\) for \(\Omega_d\in[0.05,1.2]\) at fixed perturbation; \(\sigma^*\propto\Omega_d^2\) (fitted slope \(1.97\)), while the cross-relation shows slope \(1.48\) over the scanned window (finite-drive profile corrections; see text).}
\label{fig:num_scaling}
\end{figure}

For the anisotropic potential \((\omega_x,\omega_y)=(1.0,1.6)\) the NESS is non-Gaussian and \([\mathcal{K}_s,\mathcal{K}_a]\neq0\). For \(128\) perturbations \(\Phi=\cos(k_1x+k_2y+\psi)\) with \(k_i=m_i\pi/L\), the bound \eqref{eq:quantumbound} holds in every case, with \(\max \xi_Q^2/(4\epsilon^4T^3\Sigma\sigma^*)=0.089\) (mean \(0.0098\)) [Fig.~\ref{fig:num_bound}]; the ratio is stable under grid refinement (\(0.082\), \(0.089\), \(0.090\) on \(64^2\), \(128^2\), \(256^2\) grids: the \(128^2\) and \(256^2\) values agree within \(1.5\%\), and \(\sigma^*\) is stable to better than \(0.2\%\)). Scanning \(\Omega_d\in[0.05,1.2]\) at fixed \(\Phi\), the steady-state entropy production scales as \(\sigma^*\propto\Omega_d^2\) with fitted slope \(1.97\), in agreement with the prediction. The cross-relation \(\xi_Q^2\) versus \(\sigma^*\), by contrast, shows slope \(1.48\) instead of the leading-order value \(1\) over the scanned window. This deviation is not used as a quantitative asymptotic test: at the smallest accessible drives \(\xi_Q\) becomes comparable to the discretization floor, which prevents a reliable extraction of the linear-in-\(\Omega_d\) coefficient, and at finite drive the profile corrections to the cross-scaling dominate---the bound itself is unaffected [Fig.~\ref{fig:num_scaling}].

\begin{figure}[t]
\includegraphics[width=\columnwidth]{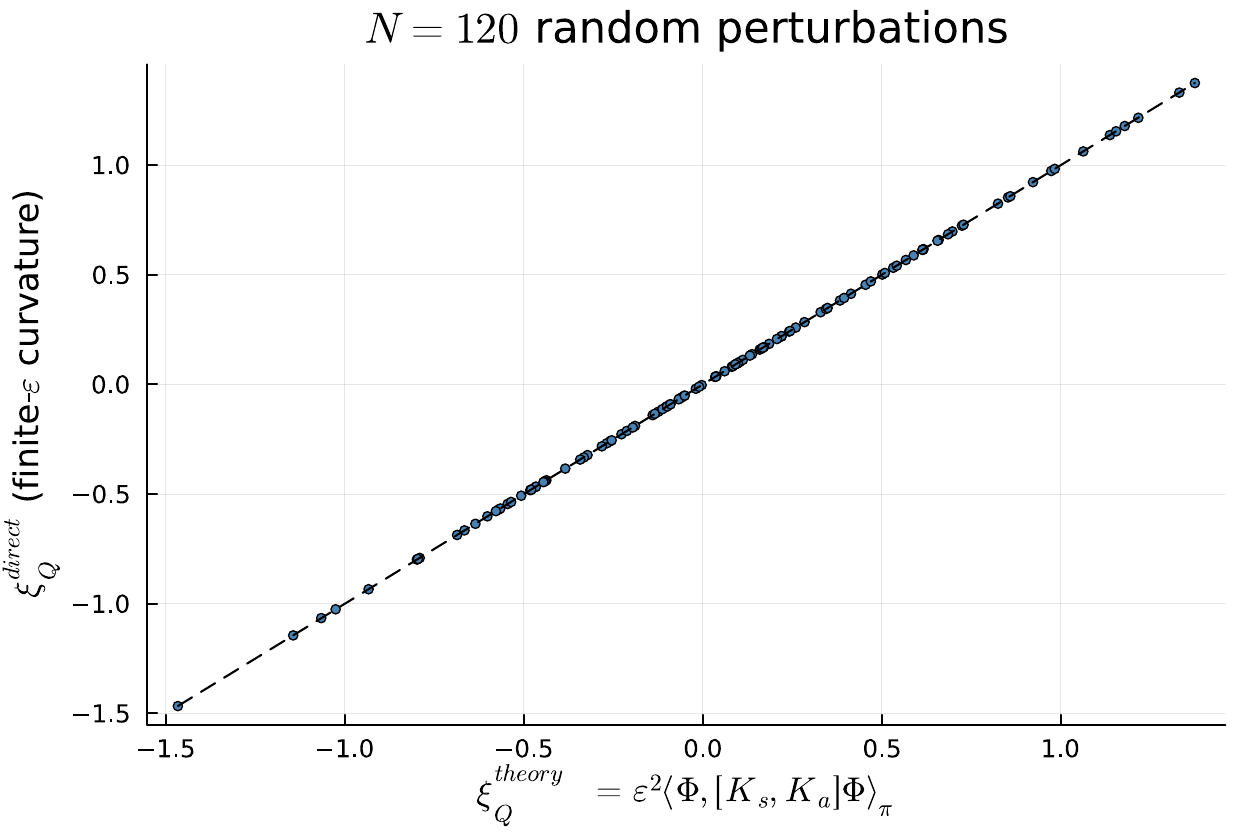}
\caption{Exact identity on the qutrit: the definition-based direct extraction \(\xi_Q^{\rm direct}\) (complex-step mixed fourth derivative) versus the algebraic expression \(\epsilon^2\langle\Phi,[\mathcal{K}_s,\mathcal{K}_a]\Phi\rangle_\pi\), for \(120\) random perturbations; all points lie on \(y=x\) (maximum relative error \(5\times10^{-4}\)).}
\label{fig:num_identity}
\end{figure}

\begin{figure}[t]
\includegraphics[width=\columnwidth]{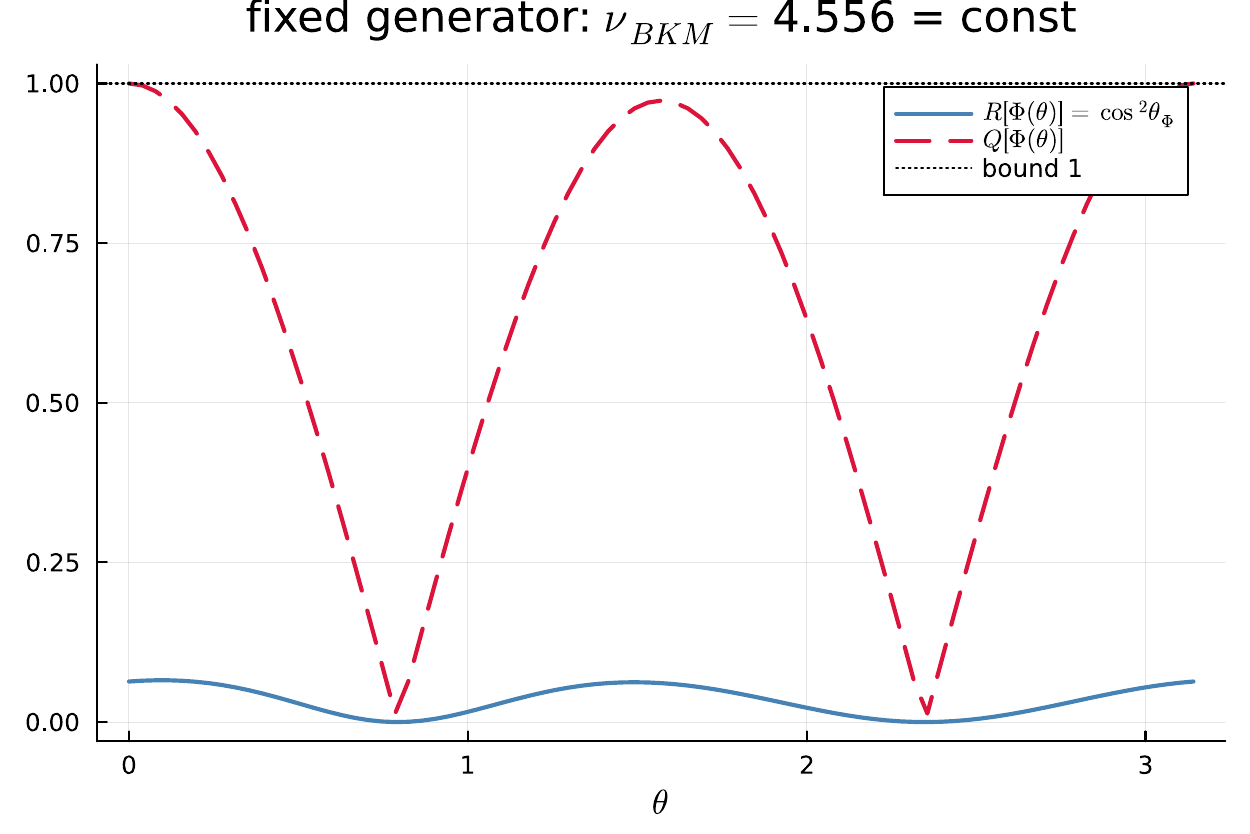}
\caption{Normalized ratio at fixed generator: \(R[\Phi(\theta)]=\cos^2\theta_\Phi\) and \(Q[\Phi(\theta)]=2|\xi_Q|/(\epsilon^2\nu_{\rm BKM}\|\Phi\|_\pi^2)\) along the family connecting the two eigendirections of \([\mathcal{K}^\ddagger,\mathcal{K}]\) with eigenvalues \(\pm\nu_{\rm BKM}\); \(Q\) sweeps \([0.013,0.9999]\) while \(\nu_{\rm BKM}=4.56\) is constant.}
\label{fig:num_visibility}
\end{figure}

To summarize, the exact quadratic curvature identity is not an algebraic artifact (Fig.~\ref{fig:num_identity}); generator-level non-normality is not equivalent to observable signal strength, which is controlled by sector alignment (Fig.~\ref{fig:num_visibility}); and in the semiclassical regime the curvature signal yields the thermodynamic lower bound (Figs.~\ref{fig:num_bound} and~\ref{fig:num_scaling}). The fully quantum link from \(\xi_Q\) to a steady-state entropy production remains the open problem.

\section{Conclusion and outlook}
\label{sec:conclusion}
We have established a local quantum information-geometric formulation of the nonequilibrium relaxation inequality for full-rank quantum Markov dynamics. The construction has three layers. First, the Kubo--Mori parametrization of perturbations turns the exact second-order expansion of the von Neumann relative entropy into the BKM inner product at all temperatures, so that the local quantum information geometry replaces the high-temperature expansion used in earlier semiclassical treatments. Second, for any quantum Markov semigroup with full-rank stationary state, the induced tangent-space generator splits canonically into a metric-dissipative part and a metric-antisymmetric transport part, and the leading \(\mathcal{O}(\epsilon^2)\) curvature correction obeys the exact identity \eqref{eq:xi_quantum}, with the universal Cauchy--Schwarz bound \eqref{eq:quantum_CS} and its collinearity saturation condition. Third, in the high-temperature, overdamped limit the identity reduces to the position-space integral relation and reproduces Auconi's entropy-production bound with the correct temperature factor.  All these  are confirmed numerically in Sec.~\ref{sec:numerics}.

The exact quadratic curvature identity and its Cauchy--Schwarz bound hold in the fully noncommutative regime. The main open step is to establish a generator-level bound of the form \(\Sigma_a(\Phi)\le C(\Phi)\,\sigma_{\rm ss}\) for a suitable quantum steady-state entropy production \(\sigma_{\rm ss}\) (for instance through a Schnakenberg-type decomposition of the generator~\cite{Spohn1978}), which would upgrade \eqref{eq:quantum_CS} to an explicit quantum entropy-production bound \(\xi_Q^2 \le 4\epsilon^4\Sigma_s(\Phi)\,C(\Phi)\,\sigma_{\rm ss}\). A further open question is whether \(\sigma_{\rm ss}\) itself controls the BKM non-normality, \(\|[\mathcal{K}^\ddagger,\mathcal{K}]\|\le F(\sigma_{\rm ss})\). Further directions are a numerical many-body test of the optimal perturbation of Eq.~\eqref{eq:ratio_bound}, time-dependent reference states \(\pi_t\) (driving protocols with a Hatano--Sasa-type correction), and settings beyond the local BKM-Markov regime, such as finite perturbations, non-Markovian memory kernels, and other monotone metrics. 

\begin{acknowledgments}
This work is supported by Yancheng Institute of Technology (xjr2024030). Z.H. is supported by the National Natural Science Foundation of China under Grant No. 12305035.

\end{acknowledgments}

\appendix
\section{Second-order expansion of the relative entropy}
\label{app:expansion}
Let \(\rho(\epsilon)=\rho^*+\epsilon\delta\) with \(\Tr\delta=0\). The relative entropy is
\(S(\epsilon)=\Tr[(\rho^*+\epsilon\delta)(\ln(\rho^*+\epsilon\delta)-\ln\rho^*)]\).
The first derivative vanishes because \(\Tr\delta=0\). For the second derivative we need the operator derivatives of the logarithm, obtained from the integral representation
\begin{equation}
\ln X = \int_0^\infty\Bigl(\frac{1}{\lambda+1}I - \frac{1}{X+\lambda I}\Bigr)d\lambda,
\end{equation}
valid for any positive operator \(X\) (see, e.g., Ref.~\cite{Hiai2011}). Differentiating with respect to \(\epsilon\), we have
\begin{equation}
\frac{d}{d\epsilon}\ln(\rho^*+\epsilon\delta)\Big|_{\epsilon=0}
= \int_0^\infty (\rho^*+\lambda)^{-1}\,\delta\,(\rho^*+\lambda)^{-1}\,d\lambda,
\label{eq:Duhamel_log}
\end{equation}
and
\begin{multline}
\frac{d^2}{d\epsilon^2}\ln(\rho^*+\epsilon\delta)\Big|_{\epsilon=0}
= -2\int_0^\infty (\rho^*+\lambda)^{-1}\,\delta\,(\rho^*+\lambda)^{-1}\\
\times\,\delta\,(\rho^*+\lambda)^{-1}\,d\lambda.
\label{eq:Duhamel_log2}
\end{multline}
Applying these to
\begin{equation}
\frac{dS}{d\epsilon} = \Tr[\delta(\ln\rho-\ln\rho^*)] + \Tr[\rho\,\tfrac{d}{d\epsilon}\ln\rho],
\end{equation}
using \(\Tr\delta=0\), and evaluating at \(\epsilon=0\), the second derivative becomes
\begin{equation}
\left.\frac{d^2S}{d\epsilon^2}\right|_{\epsilon=0}
= 2\Tr[\delta D\ln\rho^*(\delta)] + \Tr[\rho^* D^2\ln\rho^*(\delta,\delta)].
\end{equation}
With \(\rho^* = (\rho^*+\lambda)-\lambda\) the two terms combine into
\begin{multline}
\left.\frac{d^2S}{d\epsilon^2}\right|_{\epsilon=0}
= 2\int_0^\infty \lambda\,\Tr\!\bigl[(\rho^*+\lambda)^{-2}\,\delta\,(\rho^*+\lambda)^{-1}\,\delta\bigr]d\lambda\\
= \int_0^\infty \Tr\!\bigl[\delta\,(\rho^*+\lambda)^{-1}\,\delta\,(\rho^*+\lambda)^{-1}\bigr]d\lambda,
\label{eq:d2S_exact}
\end{multline}
where the last equality follows from the elementary spectral identity
\begin{multline}
\int_0^\infty\Bigl[\frac{\lambda}{(a+\lambda)^2(b+\lambda)}+\frac{\lambda}{(b+\lambda)^2(a+\lambda)}\Bigr]d\lambda\\
=\frac{\ln a-\ln b}{a-b}
=\int_0^\infty\frac{d\lambda}{(a+\lambda)(b+\lambda)},
\label{eq:spectral_identity}
\end{multline}
applied with \(a=p_m\), \(b=p_n\) and summed over \(m,n\) with weights \(|\delta_{mn}|^2\). Inserting the spectral resolution \(\rho^*=\sum_n p_n|n\rangle\langle n|\) into the last integral of~\eqref{eq:d2S_exact} gives
\begin{equation}
\int_0^\infty\frac{d\lambda}{(p_m+\lambda)(p_n+\lambda)}
= \frac{\ln p_m-\ln p_n}{p_m-p_n},
\end{equation}
which, together with \(\delta_{mn}=\langle m|\delta|n\rangle\), yields Eq.~\eqref{eq:S_exact} of the main text.

\section{Kubo--Mori inner product in the high-temperature limit}
\label{app:Kubo}
Let \(\rho^* = e^{-\beta\mathcal{H}}/Z\) with \(\beta\|\mathcal{H}\|\ll 1\). Writing 
\begin{equation}\rho^{*s}=Z^{-s}e^{-s\beta\mathcal{H}} = Z^{-s}(I - s\beta\mathcal{H}) + \mathcal{O}(\beta^2)\end{equation} 
we find
\begin{align}
\langle A,B\rangle_{\rho^*} &= \frac1Z\int_0^1 \Tr\bigl[(I - s\beta\mathcal{H})A^\dagger\nonumber\\
&\qquad\times(I - (1-s)\beta\mathcal{H})B\bigr]ds + \mathcal{O}(\beta^2) \nonumber\\
&= \frac1Z\Tr[A^\dagger B] - \frac{\beta}{2Z}\,\Tr[\{\mathcal{H},A^\dagger\}B] + \mathcal{O}(\beta^2).
\end{align}
Since \(\rho^* = Z^{-1}(I - \beta\mathcal{H}) + \mathcal{O}(\beta^2)\), this gives
\begin{equation}
\langle A,B\rangle_{\rho^*} = \Tr[\rho^* A^\dagger B] + \frac{\beta}{2Z}\Tr[[\mathcal{H},A^\dagger]B] + \mathcal{O}(\beta^2).
\end{equation}
For Hermitian \(\Phi\), i.e., \(A=B=\Phi\), the commutator correction vanishes, \(\Tr[[\mathcal{H},\Phi]\Phi] = \Tr[\mathcal{H}\Phi^2 - \Phi\mathcal{H}\Phi] = 0\) by cyclicity of the trace. This proves Eq.~\eqref{eq:KM_degenerate} of the main text. The neglected terms are bounded by \(C\beta^2\|\mathcal{H}\|^2\).

\section{Derivation of the linearized equation for \(\Phi\)}
\label{app:linearize}
Starting from \(\partial_t\delta = -\nabla\cdot(\mathbf{F}^*\delta) + T\nabla^2\delta\) and \(\delta = \rho^*\Phi\), we expand:
\begin{align}
\rho^*\partial_t\Phi &= -\nabla\cdot(\mathbf{F}^*\rho^*\Phi) + T\nabla\cdot(\rho^*\nabla\Phi + \Phi\nabla\rho^*) \nonumber\\
&= -\Phi\nabla\cdot(\mathbf{F}^*\rho^*) - \mathbf{F}^*\rho^*\cdot\nabla\Phi \nonumber\\
&\quad + T\nabla\cdot(\rho^*\nabla\Phi) + T\nabla\cdot(\Phi\nabla\rho^*).
\end{align}
Using the steady-state condition \(\nabla\cdot(\mathbf{F}^*\rho^*) = T\nabla^2\rho^*\), the first and fourth terms combine to \(T\nabla\Phi\cdot\nabla\rho^*\). Hence,
\begin{align}
\rho^*\partial_t\Phi &= T\nabla\cdot(\rho^*\nabla\Phi) + T\nabla\Phi\cdot\nabla\rho^* - \mathbf{F}^*\rho^*\cdot\nabla\Phi \nonumber\\
&= T\nabla\cdot(\rho^*\nabla\Phi) - (\mathbf{F}^*\rho^* - T\nabla\rho^*)\cdot\nabla\Phi \nonumber\\
&= T\nabla\cdot(\rho^*\nabla\Phi) - \mathbf{J}^*\!\cdot\!\nabla\Phi,
\end{align}
which, upon division by \(\rho^*\), gives \eqref{eq:Phi_lin_correct}.

\end{document}